\documentclass[aps,12pt,final,notitlepage,oneside,onecolumn,nobibnotes,nofootinbib,%
superscriptaddress,noshowpacs,centertags]{revtex4}
\usepackage{amsmath,amssymb,amsfonts,amsbsy}
\usepackage{xcolor}
\usepackage{graphicx}
\usepackage{multirow}
\usepackage[normalem]{ulem} 

\begin{document}

\title{Improved calculation of the hyperfine structure of muonic helium.}
\author{\firstname{F.~A.}~\surname{Martynenko}}
\affiliation{Samara University, Samara, Russia}
\author{\firstname{K.~A.}~\surname{Seredina}}
\affiliation{Samara University, Samara, Russia}
\author{\firstname{A.~P.}~\surname{Martynenko}}
\affiliation{Samara University, Samara, Russia}

\begin{abstract}
Using perturbation theory for the fine-structure constant $\alpha$ and electron-to-muon mass ratio, 
we calculate new contributions to hyperfine structure of the ground state of muon-electron helium atom. 
Compared to our previous results, we obtain new corrections in second-order perturbation theory. 
This calculation reduces the discrepancy between theoretical value and new experimental 
hyperfine structure measurement.
\end{abstract}


\maketitle

\section{Introduction}
\label{vv}

The three-particle muonic helium atom, consisting of an $\alpha$ particle, a muon, and an electron, is 
a bound state that, along with muonium, has been studied in quantum electrodynamics for a long time
\cite{HH,LM1,RD,Amusia,HH1,LM2,borie}. 
Measurements of the hyperfine structure of the ground state in muonic helium, performed many years ago 
in weak and strong magnetic fields, yielded the following results \cite{hughes}:
\begin{equation}
\label{eq01}
\Delta\nu^{hfs}_{exp}(\mu e ^4He)=4464.95(6)~MHz ~(13~ppm)\qquad \text{(weak field)},
\end{equation}
\begin{equation}
\label{eq02}
\Delta\nu^{hfs}_{exp}(\mu e ^4He)=4465.004(29)~MHz ~(6.5~ppm)\qquad \text{(strong field)}.
\end{equation}

The hyperfine structure of the spectrum is determined in such an atom by the interaction of spins 
of the electron and muon and differs from hyperfine structure of muonium in that in the case 
of muonic helium the electron and muon are in the Coulomb field of the nucleus.

The MuSEUM collaboration at J-PARC conducted a new series of experiments with muonic helium and muonium 
to measure the ground-state hyperfine structure \cite{muse}. The new value 
for the ground-state hyperfine splitting in muonic helium, obtained in the work \cite{muse}, is the following:
\begin{equation}
\label{eq1}
\Delta\nu^{hfs}_{exp}(\mu e ^4He)=4464.980(20)~MHz~(4.5~ppm).
\end{equation}
It is in agreement with previous measurements \cite{hughes}.  

Theoretical studies of hyperfine structure in muonic helium
have been carried out over a long period of time in various 
papers, using both the variational method and analytical perturbation theory
\cite{HH,LM1,RD,Amusia,HH1,LM2,borie,Chen,frolov,apm2008,apm2011,bekbaev,apm2022,apm2021}. Various corrections 
were calculated, and the interpretation of various effects was modified, all with the goal of obtaining 
a numerical value for the hyperfine splitting with an accuracy of 0.01 MHz, allowing for comparison 
with experiment. The difference between theoretical values in \cite{apm2008,frolov,bekbaev,apm2022}
and new experimental result is on the order of 0.5 MHz:
$\Delta\nu^{hfs}_{th}(\mu e ^4He)=4464.5042$~MHz \cite{apm2022};
$\Delta\nu^{hfs}_{th}(\mu e ^4He)=4464.55$~MHz \cite{bekbaev};
$\Delta\nu^{hfs}_{th}(\mu e ^4He)=4464.55454$~MHz \cite{frolov}.
At the same time, a number of other theoretical results are in better agreement with experiment:
$\Delta\nu^{hfs}_{th}(\mu e ^4He)=4465.1\pm 1.0$~MHz \cite{HH};
$\Delta\nu^{hfs}_{th}(\mu e ^4He)=4465.0\pm 0.3$~MHz \cite{HH1};
$\Delta\nu^{hfs}_{th}(\mu e ^4He)=4464.87\pm 0.05$~MHz \cite{Chen}.
Our previous calculations \cite{apm2008,apm2022,apm2021} of hyperfine structure in ($\mu e ^4He$) 
were performed
within the perturbation theory method proposed in \cite{LM2,LM1,LM3}.
Although \cite{apm2008,apm2022} took into account various corrections for recoil, vacuum polarization, 
and nuclear structure, the resulting discrepancy \cite{apm2008,apm2022} from the experimental value 
in \cite{muse} requires a new study of other important corrections to hyperfine structure, which 
is carried out in this paper. The results obtained within variational method indicate that 
the discrepancy may be primarily due to second-order corrections in perturbation theory (PT). In this paper, 
we calculate a number of additional contributions using perturbation theory that were not previously 
considered in \cite{apm2008,apm2022,apm2021}.

\section{General formalism}

To calculate the energy levels using the analytical perturbation theory method, we divided the Hamiltonian 
of the system into several parts, presenting the main contribution of the Coulomb interaction 
$H_0$ in the form \cite{LM1,LM2,apm2008,apm2022}:
\begin{equation}
\label{eq2}
H=H_0+\Delta H+\Delta H_{rec}+\Delta H_{vp}+\Delta H_{str}+\Delta H_{vert},~
H_0=-\frac{1}{2M_\mu}\nabla^2_\mu-\frac{1}{2M_e}
\nabla^2_e-\frac{Z\alpha}{x_\mu}-\frac{(Z-1)\alpha}{x_e},
\end{equation}
\begin{equation}
\label{eq3}
\Delta H=\frac{\alpha}{|{\bf x}_{\mu}-{\bf x}_{e}|}-\frac{\alpha}{x_e},~~~
\Delta H_{rec}=-\frac{1}{M}{\boldsymbol\nabla}_\mu\cdot{\boldsymbol\nabla}_e,
\end{equation}
where ${\bf x_\mu}$ and ${\bf x_e}$ are the radius vectors of the muon and electron relative to the nucleus, 
$Ze$ is the charge of the He nucleus. The terms $\Delta H$, $\Delta H_{vp}$, $\Delta H_{str}$, 
$\Delta H_{vert}$, and $\Delta H_{rec}$ describe the contribution of recoil effects on the electron 
and muon mass ratio, vacuum polarization, nuclear structure effects, vertex corrections, and nuclear 
recoil. The reduced masses in the electron-nucleus and muon-nucleus subsystems in the Hamiltonian $H_0$ are equal to
\begin{equation}
\label{eq4}
M_e=\frac{m_eM}{(m_e+M)},~~~M_\mu=\frac{m_\mu M}{(m_\mu+M)}.
\end{equation}
In the Hamiltonian $H_0$, it is convenient to retain the notation for the nuclear charge Z, rather than setting 
$Z=2$, as would be the case for the nucleus $^4_2He$. This allows one to correctly determine the order 
of the calculated corrections, which is related, among other things, to the factor $Z\alpha$.

In the initial approximation determined by the Hamiltonian $H_0$, the wave function of the system 
has a simple analytical form:
\begin{equation}
\label{eq5}
\Psi_0({\bf x_e},{\bf x_\mu})=\psi_{e0}({\bf x_e})\psi_{\mu 0}({\bf
x_\mu})=\frac{1}{\pi} (W_e W_\mu)^{3/2}e^{-W_\mu x_\mu}e^{-W_e x_e},~
W_\mu =Z\alpha M_\mu,~W_e=(Z-1)\alpha M_e,
\end{equation}
which makes it possible to further calculate corrections in the hyperfine structure of the spectrum 
using perturbation theory.

The hyperfine part of the Hamiltonian is determined by spin-spin interaction of the electron 
and muon in the form \cite{t4}:
\begin{equation}
\label{eq5a}
\Delta H^{hfs}=-\frac{8\pi\alpha}{3m_em_\mu}
\frac{({\boldsymbol\sigma}_e{\boldsymbol\sigma}_\mu)}{4}\delta({\bf x_\mu}-{\bf x_e}),
\end{equation}
where ${\boldsymbol\sigma}_e$ and ${\boldsymbol\sigma}_\mu$ are the electron and muon spin matrices.

Then the main contribution to hyperfine splitting 
$<{\bf s}_e{\bf s}_\mu>_{singlet}-<{\bf s}_e{\bf s}_\mu>_{triplet}=-1$ (singlet-triplet splitting) 
can be calculated analytically from the contact interaction \eqref{eq5a}:
\begin{equation}
\label{eq5b}
\Delta E^{hfs}_0=\langle\frac{8\pi\alpha(1+\kappa_\mu)}{3m_em_\mu}\delta({\bf x_\mu}-{\bf x_e})\rangle=
\frac{E_F(1+\kappa_\mu)}{\left(1+\frac{M_e}{2M_\mu}\right)^3}=4488.6167~MHz, 
~E_F=\frac{8\alpha W_e^3}{3m_em_\mu}.
\end{equation}
The formula \eqref{eq5b} includes a factor $(1+\kappa_\mu)$ with the anomalous magnetic moment 
of the muon $\kappa_\mu$. 
One-loop vertex corrections to the electron line were calculated separately in \cite{apm2022}. 
Numerical values of fundamental physical constants are taken from \cite{codata2025}.

In what follows, all calculated corrections are expressed in terms of the Fermi energy $E_F=4516.9148$~MHz. 
To obtain numerical result accurate to hundredths of a MHz, it is necessary to take into account 
second-order corrections to $E_F$ with respect to the ratio $M_e/M_\mu$ and $\alpha$.

\section{Recoil corrections in second-order perturbation theory}
\label{sec3} 

An important feature of the analytical method for calculating the hyperfine structure of muonic helium, 
compared to the variational approach, is the division of the Hamiltonian of the system with Coulomb 
interaction into several parts. This leads to a significant number of contributions appearing 
in second-order perturbation theory, which must be taken into account to ensure 
calculation 
accuracy down to hundredths of a MHz.

It is useful to recall the basic relations for calculating the recoil corrections of order 
$\alpha^4\frac{W_e}{W_\mu}$, $\alpha^4\frac{W^2_e} {W^2_\mu}\ln\frac{W_e}{W_\mu}$, and 
$\alpha^4\frac{W^2_e}{W^2_\mu}$ in the hyperfine structure
in second-order of perturbation theory. They are determined by two Hamiltonian operators 
$\Delta H$ \eqref{eq3} and $\Delta H^{hfs}$ \eqref{eq5a}.

The general expression for second-order perturbation theory
correction in hyperfine splitting is the following:
\begin{equation}
\label{eq15a}
\Delta E_1^{(2)}=2
\int\Psi^\ast({\bf x}_e,{\bf x}_\mu) \Delta H^{hfs}({\bf x}_e,{\bf x}_\mu)
\tilde G({\bf x}_e,{\bf x}_\mu;{\bf x}_e',{\bf x}_\mu ')
\Delta H({\bf x}_e',{\bf x'}_\mu)\Psi({\bf x}_e',{\bf x}_\mu ')d{\bf x}_e
d{\bf x}_\mu d{\bf x}_e' d{\bf x}_\mu ',
\end{equation}
where the superscript $(2)$ here and below denotes second-order perturbation theory contribution. 
One integration over coordinates can be performed in \eqref{eq15a} using the Dirac $\delta$-function. 
The expression \eqref{eq15a} contains the reduced Coulomb Green's function:
\begin{equation}
\label{eq16}
\tilde G({\bf x}_e,{\bf x}_\mu;{\bf x'}_e,{\bf x'}_\mu)=\sum_{n,n'\not =0}
\frac{\psi_{\mu n}({\bf x}_\mu)
\psi_{en'}({\bf x}_e)\psi^\ast_{\mu n}({\bf x'}_\mu)\psi^\ast_{en'}({\bf x'}_e)}{E_{\mu 0}+
E_{e0}-E_{\mu n}-E_{en'}},
\end{equation}
where $E_{e0}=-M_e(Z-1)^2\alpha^2/2$, $E_{\mu 0}=-M_\mu Z^2\alpha^2/2$ are the Coulomb energies 
of the electron and
muon, which are determined by the initial approximation of the Hamiltonian $H_0$ \eqref{eq2}.

When calculating the correction \eqref{eq15a}, it is convenient to split the sum over the muon states in equation 
\eqref{eq16} into two parts with $n=0$ and $n\not=0$. The general expression for the first contribution is equal to
\begin{equation}
\label{eq17}
\Delta E_1^{(2)}(n=0)=\frac{4\pi \alpha}{3}\frac{g_eg_\mu}{m_em_\mu}\int|\psi_{\mu 0}({\bf x}_3)|^2
\psi^\ast_{e0}({\bf x}_3)\sum_{n'\not=0}^\infty
\frac{\psi_{en'}({\bf x}_3)\psi^\ast_{en'}({\bf x}_1)}{E_{e0}-E_{en'}}V_\mu({\bf x}_1)
\psi_{e0}({\bf x}_1)d{\bf x}_1d{\bf x}_3,\
\end{equation}
\begin{equation}
\label{eq18}
V_\mu({\bf x}_1)=\int\psi^\ast_{\mu 0}({\bf x}_2)\left[\frac{\alpha}{|{\bf x}_2-{\bf x}_1|}-\frac{\alpha}{x_1}\right]\psi_{\mu 0}({\bf x}_2)d{\bf x}_2=
-\frac{\alpha}{x_1}(1+W_\mu x_1)e^{-2W_\mu x_1}.
\end{equation}
For the reduced Coulomb Green's function of the electron included in \eqref{eq17}, a compact expression 
was obtained in \cite{Hameka} in the form:
\begin{equation}
\label{eq19}
\tilde G_e({\bf x}_1,{\bf x}_3)=\sum_{n\not =0}^\infty\frac{\psi_{en}({\bf
x}_3) \psi_{en}^\ast({\bf x}_1)}{E_{e0}-E_{en}}=-\frac{W_e M_e}{\pi}e^{-W_e(x_1+x_3)}
\Biggl[\frac{1}{2W_e x_>}-
\end{equation}
\begin{displaymath}
-\ln(2W_e x_>)-\ln(2W_e x_<)+Ei(2W_e x_<)+
\frac{7}{2}-2C-W_e(x_1+x_3)+\frac{1-e^{2W_e x_<}}{2W_e x_<}\Biggr],
\end{displaymath}
where $x_<=\min(x_1,x_3)$, $x_>=\max(x_1,x_3)$, $C=0.577216\ldots$ is Euler's constant, and $Ei(x)$ 
is the exponential integral function. Then, integration over the coordinates in \eqref{eq17}
can be performed analytically, and the obtained result can be represented as an expansion 
in the parameter $W_e/W_\mu$:
\begin{equation}
\label{eq20}
\Delta E_1^{(2)}(n=0)=E_F(1+\kappa_\mu)\left[\frac{11}{8}\frac{W_e}{W_\mu}-\frac{1}{16}
\frac{W_e^2}{W_\mu^2}\left(64\ln\frac{W_e}{W_\mu}+64\ln 2+7\right)\right].
\end{equation}
The excited states of the muon $(n\not=0)$ give second part of the contribution to hyperfine 
structure in the form:
\begin{equation}
\label{eq21}
\Delta E_1^{(2)}(n\not=0)=\frac{4\pi\alpha}{3}\frac{g_eg_\mu}{m_em_\mu}\int\psi^\ast_{\mu 0}({\bf x}_3)
\psi^\ast_{e 0}({\bf x}_3)
\sum_{n\not=0}\psi_{\mu n}({\bf x}_3)\psi^\ast_{\mu n}({\bf x}_2)G_e({\bf x}_3,{\bf x}_1,z)\times
\end{equation}
\begin{displaymath}
\left[\frac{\alpha}{|{\bf x}_2-{\bf x}_1|}-\frac{\alpha}{x_1}\right]\psi_{\mu 0}({\bf x}_2)
\psi_{e 0}({\bf x}_1)d{\bf x}_1 d{\bf x}_2d{\bf x}_3.
\end{displaymath}
where the electron Green's function is introduced:
\begin{equation}
\label{eq22}
G_e({\bf x}_3,{\bf x}_1,z)=\sum_{n'=0}^\infty\frac{\psi_{en'}({\bf x}_3)
\psi^\ast_{en'}({\bf x}_1)}{z-E_{en'}}=
\sum_{n'=0}^\infty\frac{\psi_{en'}({\bf x}_3)\psi^\ast_{en'}({\bf x}_1)}
{E_{\mu 0}+E_{e0}-E_{\mu n}-E_{en'}}.
\end{equation}
The term $(-\alpha/x_1)$ in \eqref{eq21} does not contribute due to the orthogonality of muon wave 
functions. To perform further analytical integration in \eqref{eq21}, we approximately replace $G_e$ 
with the free Green's function \cite{LM1,LM2}:
\begin{equation}
\label{eq23}
G_e({\bf x}_3,{\bf x}_1,E_{\mu 0}+E_{e0}-E_{\mu n})\to G_{e0}({\bf x}_3-{\bf x}_1,E_{\mu 0}+E_{e0}-E_{\mu n})=
-\frac{M_e}{2\pi}\frac{e^{-\beta|{\bf x}_3-{\bf x}_1|}}{|{\bf x}_3-{\bf x}_1|},
\end{equation}
where $\beta=\sqrt{2M_e(E_{\mu n}-E_{e0}-E_{\mu 0})}$. In addition, we employ another approximation 
by replacing the electron wave functions in \eqref{eq21} with their values at zero, $\psi_{e0}(0)$.
Terms omitted in this approximation can give 
a second-order contribution with respect to $\frac{W_e}{W_\mu}$. The results of numerical integration 
in \cite{LM1} with the exact electron Green's function in the case of muonic helium show that the terms 
used in the approximation \eqref{eq23} are numerically small.

After these approximations, integration over the coordinate ${\bf x}_1$ gives the following result:
\begin{equation}
\label{eq24}
\int\frac{e^{-\beta|{\bf x}_3-{\bf x}_1|}}{|{\bf x}_3-{\bf x}_1|}
\frac{d{\bf x}_1}{|{\bf x}_2-{\bf x}_1|}=
4\pi\left[\frac{1}{\beta}-\frac{1}{2}|{\bf x}_3-{\bf x}_2|+\frac{1}{6}\beta|{\bf x}_3-{\bf x}_2|^2-
\frac{\beta^2}{24}|{\bf x}_3-{\bf x}_2|^3+\ldots\right],
\end{equation}
where an expansion of $e^{-\beta|{\bf x}_2-{\bf x}_3|}$ in $\beta|{\bf x}_2-{\bf x}_3|$ is equivalent 
to the expansion in powers of $\sqrt{W_e/W_\mu}$. The first term $\beta^{-1}$ of the expansion in \eqref{eq24} 
does not contribute to \eqref{eq21}. The second term of the expansion in \eqref{eq24} gives a leading-order 
contribution in $\sqrt{W_e/W_\mu}$: $-E_F\frac{35W_e}{8W_\mu}$. To improve the accuracy of the result, 
we also consider the third term on the right-hand side of \eqref{eq24}, which leads to the following integral:
\begin{equation}
\label{eq25}
\int \psi^\ast_{\mu 0}({\bf x}_3)\sum_{n}\sqrt{2M_e(E_{\mu n}-E_{\mu 0})}
\psi_{\mu n}({\bf x}_3)\psi^\ast_{\mu n}({\bf x}_2)({\bf x}_2\cdot{\bf x}_3)
\psi_{\mu 0}({\bf x}_2)d{\bf x}_2d{\bf x}_3=\sqrt{\frac{2W_e}{W_\mu^3}}S_{\frac{1}{2}},
\end{equation}
where the value is entered
\begin{equation}
\label{eq26}
S_{1/2}=\sum_{n}\left(\frac{E_{\mu n}-E_{\mu 0}}{R_\mu}\right)^{1/2}|\langle\mu 0|
\frac{\bf x}{a_\mu}|\mu n\rangle|^2, ~~~R_\mu=2M_\mu\alpha^2.
\end{equation}

The contribution to equation \eqref{eq26} is made by the matrix elements for the discrete and continuous states, 
which are presented in the book \cite{HBES}. The numerical contributions of discrete and continuous 
states to equation \eqref{eq26} are as follows:
\begin{equation}
\label{eq27}
S_{\frac{1}{2}}^{d}=\sum_{n}\frac{2^{8}n^6(n-1)^{2n-\frac{9}{2}}}{(n+1)^{2n+\frac{9}{2}}}=1.90695...,
\end{equation}
\begin{equation}
\label{eq27c}
S_{\frac{1}{2}}^{c}=\int_0^\infty \frac{2^8kdk}{(k^2+1)^{9/2}(1-e^{-\frac{2\pi}{k}})}
\left|\left(\frac{1+ik}{1-ik}\right)^{i/k}\right|=1.03111...  .
\end{equation}

By adding the recoil corrections \eqref{eq20} and \eqref{eq21}
in second-order perturbation theory, we obtain the total recoil correction of order $\alpha^4$ 
in the following form:
\begin{equation}
\label{eq28}
\Delta E_1^{(2)}=E_F (1+\kappa_\mu)\left[-3\frac{W_e}{W_\mu}+
\frac{231W_e^2}{32W_\mu^2}-
\frac{4W_e^2}{W_\mu^2}\ln\frac{2W_e}{W_\mu}+\frac{4W_e}{3W_\mu}\sqrt{\frac{2W_e}{W_\mu}}S_{1/2}\right],
\end{equation}
where $S_{1/2} = S_{\frac{1}{2}}^{d} + S_{\frac{1}{2}}^{c}$ given by \eqref{eq27} and \eqref{eq27c}.
The fourth term in expansion \eqref{eq24} makes no contribution to the energy spectrum.

\begin{table}[htbp]
\label{t1}
\caption{Contributions to hyperfine splitting of the ground state in muonic helium atom.}
\label{tb1}
\begin{tabular}{|| c | c ||}  \hline
Contribution to  $\Delta E^{hfs}$                     & Numerical value in MHz, reference           \\  \hline
Contribution of order $\alpha^4$     & 4488.6167, ~\eqref{eq5b}     \\  \hline
Recoil correction of order $\alpha^4\frac{W_e}{W_\mu}$, $\alpha^4\frac{W^2_e}{W^2_\mu}$& -29.7371, ~\eqref{eq28}, \cite{apm2022}    \\ \hline                                                                            
Correction of second order PT with the Breit potential & 0.4248,~\eqref{eq31}, \eqref{eq33a}, \eqref{eq35}, \eqref{eq35d}  \\   \hline                                                                            
One-loop VP correction in $1 \gamma $ interaction             &0.0359,~\cite{apm2008,apm2022}         \\ \hline                          
One-loop VP correction in $\mu N$ interaction            &0.0484,~\cite{apm2008,apm2022}                  \\
in second-order PT  &        \\    \hline  
One-loop VP correction  in $\mu e$ interaction            &-0.1012, ~\cite{apm2008,apm2022}     \\
in second-order PT                  &                             \\ \hline    
One-loop VP correction  in $ e N$ interaction                  &0.0756, ~\cite{apm2008,apm2022}     \\
in second-order PT           &                           \\ \hline 
One-loop VP correction with $\Delta H$ potential               &-0.0732, ~\cite{apm2008,apm2022}     \\ 
in second-order PT                &                               \\ \hline                                       
Nuclear structure correction in $1 \gamma$ interaction         &   0                  \\ \hline                                                                           
Nuclear structure correction in $2 \gamma$ interaction    & 0                       \\ \hline                                                                               
Nuclear structure correction in second order PT          & -0.0097, ~\cite{apm2022}       \\ \hline 
Nuclear recoil correction from $\Delta H_{rec}$          & 0.0809,  ~\cite{apm2022}     \\ \hline  
Electron vertex correction of order $\alpha^5$              & 5.1774, ~\cite{apm2022}    \\
from $1\gamma$ interaction                &                          \\ \hline                                        
Electron vertex correction of order $\alpha^5$             & -0.0206, ~\cite{apm2022}       \\
in second order PT                  &                              \\ \hline                                        
Recoil correction from $2\gamma$ interaction     &  0.8056, ~\eqref{eq37}, \cite{arnowitt,Chen}    \\ \hline                               
Radiative correction of order $\alpha^6$                  & -0.4346, ~\eqref{eq38}, \cite{LM1,borie,Chen,brodsky,kroll,karplus,salpeter}\\ \hline                                                                            
Summary contribution                        & 4464.8889                       \\ \hline
\end{tabular}
\end{table}

\section{Second-order perturbation theory corrections from the Breit Hamiltonian}
\label{sec4} 

Let us consider other contributions of second-order perturbation theory, which are determined by potentials 
from the Breit Hamiltonian. One such contribution is determined by a second-order perturbation theory 
formula with two hyperfine perturbation potentials:
\begin{multline}
\label{eq29}
\Delta E_2^{(2)}=
\int \Psi^\ast({\bf x}_e,{\bf x}_\mu)\Delta H^{hfs}({\bf x}_e,{\bf x}_\mu)\tilde G({\bf x}_e,{\bf x}_\mu;{\bf x'}_e,{\bf x'}_\mu) \times \\
\Delta H^{hfs}({\bf x'}_e,{\bf x'}_\mu) \Psi({\bf x'}_e,{\bf x'}_\mu)d{\bf x'}_e d{\bf x'}_\mu d{\bf x}_e d{\bf x}_\mu.
\end{multline}

Substituting into \eqref{eq29} the wave functions and the Green's function and performing two integrations 
using $\delta$-functions, we obtain:
\begin{equation}
\label{eq30}
\Delta E_2^{(2)}=\frac{64\alpha^2W_e^3 W_\mu^3}{9m_e^2m_\mu^2}\int 
e^{-W_\mu x_e(1+\frac{W_e}{W_\mu})}e^{-W_\mu x'_e(1+\frac{W_e}{W_\mu})}\times
\end{equation}
\begin{displaymath}
\sum_{n,n'\not =0}\frac{\psi_{\mu n}({\bf x}_e)
\psi_{en'}({\bf x}_e)\psi^\ast_{\mu n}({\bf x'}_e)\psi^\ast_{en'}({\bf x'}_e)}{E_{\mu 0}+
E_{e0}-E_{\mu n}-E_{en'}} d{\bf x'}_e d{\bf x}_e.
\end{displaymath}

As before, it is convenient to separate the two contributions with $n=0$ and $n\not=0$ into a sum. 
Calculating the matrix elements for the contribution with $n=0$, we obtain:
\begin{equation}
\label{eq31}
\Delta E_2^{(2)}(n=0)=-E_F\frac{4\alpha^2 Z M_e M_\mu }{3 m_e m_\mu }
\Bigl[ \frac{5}{4}-\frac{W_e}{W_\mu}\bigl(4\ln\frac{2W_e}{W_\mu}+\frac{7}{8}\bigr)\Bigr]=-0.8111~MHz.
\end{equation}

The second part of the correction \eqref{eq30} with $n\not=0$ is determined by the formula:
\begin{equation}
\label{eq32}
\Delta E_2^{(2)}(n\not=0)=\frac{64\alpha^2W_e^3 W_\mu^3}{9m_e^2m_\mu^2}\sum_{n\not =0}
\int e^{-W_\mu x_e(1+\frac{W_e}{W_\mu})}\psi_{\mu n}({\bf x}_e)d{\bf x}_e\times
\end{equation}
\begin{displaymath}
\int e^{-W_\mu x'_e(1+\frac{W_e}{W_\mu})} \psi_{\mu n}({\bf x'}_e)d{\bf x'}_e
G_e({\bf x}_e,{\bf x'}_e,\beta_n).
\end{displaymath}
This expression contains the electron's Coulomb Green's function $G_e({\bf x}_e,{\bf x'}_e,\beta_n)$, 
for which, in leading order in $M_e/M_\mu$, the substitution \eqref{eq23} can again be used. 
After subtracting the contribution of the iterative term of the quasipotential 
$ <V_{1\gamma}\times G^f\times V_{1\gamma}>$,
where $G^f$ is the free Green's function, this term cancels out in the energy spectrum.

Let us now consider second-order perturbation theory corrections, which are determined 
by the paired contact terms of the Breit potential:
\begin{equation}
\label{eq33}
\Delta H^{\mu e}_{cont}=-\frac{\pi\alpha}{2}\left(\frac{1}{m_e^2}+\frac{1}{m_\mu^2}\right)
\delta({\bf x}_e-{\bf x}_\mu),~~~\Delta H^{e\alpha}_{cont}=\frac{\pi Z\alpha}{2}
\left(\frac{1}{m_e^2}+\frac{1}{m_\alpha^2}\right)\delta({\bf x}_e).
\end{equation}

For these perturbation potentials, the contribution of muonic excited states $n\not=0$ will be zero, 
as for the previous correction. For intermediate muon states with $n=0$, the contact terms 
\eqref{eq33} give a nonzero contribution. For the muon-electron contact interaction, it is equal to
\begin{equation}
\label{eq33a}
\Delta E^{\mu e}_{cont}=-2\langle \Delta H^{hfs}\cdot\tilde G\cdot\Delta H^{\mu e}_{cont}\rangle=
\end{equation}
\begin{displaymath}
E_F\frac{\alpha^2(Z-1)M_e^2}{4m_e^2}\left(\frac{W_\mu}{W_e}\right)^6\int_0^\infty x_e^2dx_e\int_0^\infty
{x'}_e^2 dx'_e e^{-x_e(1+\frac{W_\mu}{W_e})}e^{-{x'}_e(1+\frac{W_\mu}{W_e})}\times
\end{displaymath}
\begin{displaymath}
\Biggl[-\ln(x_>)-\ln(x_<)+Ei( x_<)+
\frac{7}{2}-2C-\frac{(x_1+x_3)}{2}+\frac{1-e^{ x_<}}{ x_<}\Biggr]=
\end{displaymath}
\begin{displaymath}
E_F\frac{\alpha^2(Z-1) M_e^2}{4 m_e^2}\left(\frac{43}{8}-4\ln\frac{2W_e}{W_\mu}\right)=
1.5985~MHz.
\end{displaymath}
 
The contribution of the electron-nucleus contact interaction \eqref{eq33} must be considered together with 
the relativistic correction for electron motion:
\begin{equation}
\label{eq34a}
\Delta H'_e=\frac{\pi Z\alpha}{2m_e^2}\delta({\bf x}_e)-\frac{{\bf p}^4_e}{8m_e^3}.
\end{equation}

Applying the operator $\Delta H'_e$ to the electron wave function $\psi_{e0}({\bf x}_e)$, we obtain:
\begin{equation}
\label{eq34b}
\Delta H'_e\psi_{e0}({\bf x}_e)=\left[\frac{\pi Z\alpha}{2m_e^2}\delta({\bf x}_e)-
\frac{\pi (Z-1)\alpha M_e}{m_e^3}\delta({\bf x}_e)-\frac{W_e^4}{8m_e^3}\left(1-\frac{4}{W_e x_e}\right)
\right]\psi_{e0}({\bf x}_e).
\end{equation}
The first two terms in the square brackets approximately cancel out, so the contribution 
comes from the third term in \eqref{eq34b} in following integral form:
\begin{equation}
\label{eq34}
\Delta E^{\alpha e}_{cont+rel}=-\frac{16\pi\alpha}{3m_e m_\mu}\int \psi_{e0}({\bf x}_e)|\psi_{\mu 0}({\bf x}_e)|^2
\tilde G_e({\bf x}_e,{\bf x'}_e)\Delta H'_e({\bf x'}_e)\psi_{e0}({\bf x'}_e) d{\bf x'}_e d{\bf x}_e.
\end{equation}

Using further the expression for the reduced Coulomb Green's function \eqref{eq19} 
we can make an analytical integration in \eqref{eq34} and obtain the following contribution:
\begin{equation}
\label{eq35}
\Delta E^{\alpha e}_{cont}=-E_F\frac{3\alpha^2 M_e^3}{2m_e^3}\left(1-\frac{W_e}{W_\mu}\right)
=-0.3598~MHz.
\end{equation}

For the muon, there is also a contribution to the Hamiltonian analogous to expression \eqref{eq34a}, 
consisting of a relativistic correction and a contact interaction, as follows:
\begin{equation}
\label{eq35b}
\Delta H'_\mu=\frac{\pi Z\alpha}{2m_\mu^2}\delta({\bf x}_\mu)-\frac{{\bf p}^4_\mu}{8m_\mu^3}.
\end{equation}

When this perturbation is used in the second order of perturbation theory, the excited electronic states 
do not contribute. Following transformations analogous to those performed for an electron, 
this contribution can be expressed in integral form:
\begin{equation}
\label{eq35c}
\Delta E^{\alpha \mu}_{cont+rel}=-\frac{8\pi\alpha W_\mu^3}{3m_e m_\mu^4}\int |\psi_{e0}({\bf x}_e)|^2
\psi_{\mu 0}({\bf x}_e)
\tilde G_\mu({\bf x}_e,{\bf x}_\mu) \frac{1}{x_\mu}\psi_{\mu 0}({\bf x}_\mu) d{\bf x}_\mu d{\bf x}_e,
\end{equation}
where the reduced Coulomb Green's function for the muon $\tilde G_\mu({\bf x}_e,{\bf x}_\mu)$
is defined by an expression analogous to \eqref{eq19}.

By performing analytical integration over two coordinates and using an expansion in terms of a small 
parameter $W_e/W_\mu$, we obtain the following formula for this contribution:
\begin{equation}
\label{eq35d}
\Delta E^{\alpha \mu}_{cont}=-E_F\frac{W_eW_\mu M_\mu}{2m_\mu^3}\left(3-12\frac{W_e}{W_\mu}+30
\frac{W_e^2}{W_\mu^2}\right)=-0.0033~MHz.
\end{equation}

In second-order perturbation theory, there are also relativistic recoil corrections, which are determined 
by the Breit Hamiltonian for pairwise interactions. Thus, in the case of electron-nucleus interaction, 
the corresponding interaction potential has the form:
\begin{equation}
\label{eq35a}
\Delta H^{B}_{\alpha e,rec}=-\frac{Z\alpha}{2m_em_N x_e}\left({\bf p}^2_e-\frac{d^2}{dx_e^2}\right).
\end{equation}

The only contribution comes from the intermediate state of the muon with $n=0$:
\begin{equation}
\label{eq35bb}
\Delta E^{hfs}_{\alpha e,rec}=\frac{8\pi\alpha}{3m_e^2m_\mu m_N}\int |\psi_{\mu 0}({\bf x}_3)|^2
\psi_{e0}({\bf x}_3)
\tilde G({\bf x}_3,{\bf x}_e)\frac{1}{x_e}\left({\bf p}^2_e-\frac{d^2}{dx_e^2}\right)\psi_{e0}({\bf x}_e)
d{\bf x}_e d{\bf x}_3=
\end{equation}
\begin{displaymath}
E_F\frac{Z\alpha^2 M_e^2}{2m_em_N}\left[\frac{3}{2}-2\ln\frac{W_e}{W_\mu}+2\frac{W_e}{W_\mu}
\left(3\ln\frac{W_e}{W_\mu}-2\right)\right]=0.0004~ MHz.
\end{displaymath}

The total correction from the second-order perturbation theory terms \eqref{eq31}, \eqref{eq33a}, 
\eqref{eq35}, \eqref{eq35bb} is presented in Table~\ref{tb1} as a separate line.

\section{Conclusion}

Our previous study of the hyperfine structure of muonic helium was performed within 
the analytical perturbation theory approach \cite{LM1,LM2,apm2022}. It included the calculation of various 
corrections 
connected with vacuum polarization, nuclear structure, and vertex electron corrections. The complete theoretical 
result in \cite{apm2022} was in good agreement with the calculations within the framework of the variational 
method \cite{bekbaev}. At the same time, the discrepancy between the theoretical calculations and the 
new experimental 
value \eqref{eq1} is about 0.5 MHz. In this work, a number of new corrections are calculated in second-order 
perturbation theory with the Breit Hamiltonian, which are obtained in analytical form \eqref{eq31}, 
\eqref{eq33a}, \eqref{eq35}. The total numerical value of the obtained correction is 0.4248 MHz. As a result, 
the new theoretical value of hyperfine splitting in muonic helium is 4464.8889  MHz, which differs from 
the experimental value \eqref{eq1} by 0.0911 MHz. Thus, the calculated contributions to hyperfine 
splitting (HFS) lead to significantly better agreement with the experimental data. The remaining difference 
of 0.0911 MHz between the theoretical value and the experimental result is due to the approximations made in 
\cite{apm2022} when calculating various corrections in second-order perturbation theory.

In this paper, as in \cite{apm2022}, numerical results are presented for definiteness with an accuracy 
of four decimal places. The theoretical uncertainty can be estimated based on the approximations used 
in \cite{apm2022} to calculate the corrections in second-order perturbation theory, when the exact 
reduced Green's function was replaced by the free Green's function. This uncertainty does not exceed 
$0.1\div 0.2$ MHz.

Table~\ref{tb1} includes first- and second-order perturbation theory corrections that describe known effects 
in bound-state quantum electrodynamics: vacuum polarization effects, nuclear structure effects, recoil 
effects, and vertex corrections. A more detailed description of the calculation of these corrections and, 
accordingly, the results in Table~\ref{tb1} is given in \cite{apm2022}. Along with the corrections we calculated, 
Table~\ref{tb1} includes several results obtained in calculating the hyperfine structure of muonium. Since 
the muonic helium atom can be represented as a two-particle system consisting of an electron 
and a pseudonucleus (muon-alpha particle), the analytical results can be used for the corrections 
calculated for hydrogen-like systems. From the numerous tables with results in the work \cite{egs}, 
one can single out those corrections that in muonic helium must be taken into account first of all, 
since they make a numerically significant contribution.

These corrections include 
the recoil correction from the two-photon exchange diagrams 
for the electron-muon system \cite{Chen,chen1,arnowitt}:
\begin{equation}
\label{eq37}
\Delta E^{2\gamma}_{rec}=E_F\frac{3\alpha}{\pi}\frac{m_e m_\mu}{m_\mu^2-m_e^2}\ln\frac{m_\mu}{m_e}.
\end{equation}

Another numerically important contribution is determined by the two-photon exchange amplitudes 
with radiative corrections to the electron line \cite{brodsky,kroll,karplus,salpeter}:
\begin{equation}
\label{eq38}
\Delta E^{rad}=E_F (Z-1)\alpha^2 \left(\ln 2-\frac{5}{2}\right) .
\end{equation}

We emphasize that these corrections are related to two-photon exchanges, so the consideration of various 
second-order contributions of perturbation theory performed in this paper seems to be a necessary 
addition to the calculations previously performed in \cite{apm2022}.

\begin{acknowledgments}
This work is supported by the Russian Science Foundation (grant no. RSF 25-72-00029) (F.A.M.).
\end{acknowledgments}

\end{document}